# Microfluidics platform for polymorph screening directly from powder

Guillem Peybernès[1,2], Romain Grossier[1], Frédéric Villard[2], Philippe Letellier[2], Nadine Candoni[1], Stéphane Veesler[1*],

[1]CNRS, Aix-Marseille University, CINaM (Centre Interdisciplinaire de Nanosciences de Marseille), Campus de Luminy, Case 913, F-13288 Marseille Cedex 09, France,
[2]Technologie Servier, 27 Rue EugèneVignat,45000 Orléans, France
*veesler@cinam.univ-mrs.fr

## ABSTRACT

We describe a microfluidic platform for solid phase screening using extremely small quantities of raw materials. Based on our previous set-up for solubility measurement that generates saturated solutions directly from powder, the platform requires no solution in excess of that used for the droplet-crystallization experiment. The set-up is compatible with most solvents and molecules without using surfactant.
Using this microfluidic platform, we first measured the solubility of Sulfathiazole in water, isopropanol, and acetonitrile. Second, we performed a polymorph screening of Sulfathiazole using as little as 30 mg of raw material, for numerous identical cooling crystallization experiments from 80 to 10°C. In the experiments presented, we obtained the 3 usual polymorphs of Sulfathiazole. We show that this economical approach yields reliable information on the probability of nucleation of a given polymorph, useful in pharmaceutical development.

## INTRODUCTION

Polymorphism is the ability of a molecule to exist in more than one crystalline structure. As a result, polymorphs have different physicochemical properties that can profoundly influence the bioavailability, manufacturability, purification, stability and solid characteristics of an active pharmaceutical ingredient (API)[1]. Thus, discovering a new polymorph can delay or extend marketing, as with Zantac[2] and Norvir[3]. This makes polymorph or general solid phase screening of APIs essential[4] [5] in pharmaceutical development. The current empirical approach involves screening the crystallization parameters such as supersaturation, temperature, solvent composition, impurities, cooling rates and hydrodynamics[1]. However, this is costly in terms of raw materials. For instance, Morissete et al.[4] report a range of 1–10 mg of drug per experiment. Moreover, as nucleation is stochastic, it requires numerous experiments to obtain reliable data[6-8], increasing the consumption of raw materials. In an attempt to develop methods that use smaller amounts of materials, small-volume experimental devices have been proposed. For example, manipulating fluids at submillimeter scale was first proposed for protein crystallization with high-throughput robotic techniques[9, 10], and was extended to small molecules with arrays of microdroplets of identical composition[8] and a lab-on-chip approach[11]. Lab-on-chip or microfluidics platforms/devices for crystallization screening have recently received great interest in the literature[12-14]. However, generation and manipulation of super- and saturated solutions is often an issue. The usual method involves preparing a stock solution by adding an excess of powder to a given volume of solvent at a given temperature. This suspension is stirred for 24 to 48 hours at the desired temperature to reach the thermodynamic equilibrium. The suspension is decanted and the supernatant is collected and filtered. The experimental difficulty with this protocol is to maintain the temperature constant during all the steps, to avoid unwanted crystallization. Moreover in microfluidics experiments, even if the crystallization experiment consumes only a few µL, stock solutions of at least hundreds of µL are required.

Our aim was, therefore, to develop a microfluidics platform for solid phase screening that eliminates the stock solution. We started from the set-up we previously developed for solubility measurement based on the generation of saturated solutions directly from powder[15]. There is no need to prepare more solution than what is required for the droplet-crystallization experiment, and the set-up is compatible with most solvents and molecules without using surfactant.

In this paper, we describe our microfluidics platform and how it is used to perform a polymorph screening of Sulfathiazole in water, isopropanol and acetonitrile using extremely small quantities of material, as little as 30 mg, for cooling crystallization experiments from 80 to 10°C. This microfluidics approach was used to perform numerous identical experiments yielding consistent data on polymorph nucleation.

## EXPERIMENTAL SECTION

### Testing Molecules

Sulfathiazole (4-Amino-N-(2-thiazolyl)benzenesulfonamide) was purchased from Fluka Analytical (MKBQ0002V, batch 1001670734). Sulfathiazole is known to exhibit five polymorphic forms: I, II, III, IV, and V[16, 17]. It is generally accepted that relative thermodynamic stabilities follow the order of the densities of the structure, i.e., III $\cong$ IV> II > I. The form III is supposed to be the stable form, but in a paper Blagden et al.[18] stated that form IV is the most stable phase. Form I is the most metastable at room temperature; form V has only been crystallized from boiling water and is reputedly very unstable, following a fast solution-mediated-phase transition to form I[8]. The commercial sulfathiazole we purchased is the stable form III as determined by X-ray diffraction analysis.

GPL Krytox® fluorinated oils were purchased from Dupont, FMS and FC70 from Hampton Research. Solvents are analytical grade.

### Microfluidics Set-up

All the set-ups presented in this paper are PEEK (polyether ether ketone) devices based on HPLC techniques (IDEX Health and Science). T-junctions, precolumn filters, shut-off valves, and PFA (Perfluoroalkoxy alkane) tubing (1mm ID), which are resistant to many solvents, render the device applicable to mineral, organic and biological materials. The solutions are loaded using separate syringes and a programmable syringe pump (neMESYS, cetoni GmbH) controls the flow rates of the different fluids.

#### *Solubility measurement*

Sulfathiazole solubility was measured for isopropanol, acetonitrile and water at temperatures between 20 and 80 °C with the microfluidics set-up and experimental procedure presented previously[15]. This set-up is used to measure in 4 hours, directly from powder (as little as 30mg), a temperature solubility curve (5 to 7 points) in a solvent, with approximately 5% error.

#### *On-line preparation of saturated solutions without pre-knowledge of solubility*

To overcome the risk of unwanted crystallization during handling and save materials, we have improved a microfluidics technique, that we previously developed for solubility measurement[15]. In this extraction set-up (i.e. dissolution of powder into solvent), the solvent flows through the powder bed blocked by a filter. In practice, 30mg of powder is placed in a 1mm inner diameter (ID) tube connected to a precolumn filter of 0.5µm maintained at constant temperature in an incubator (point **a** on figs.1 and S1). At the outlet of the filter, due to the dissolution of the powder the solution is at saturation.

*Plug factory: generation of saturated droplets*
Droplets are generated by cross-flowing in a T-junction (point **b** on fig.1). The main channel contains the continuous phase, an inert fluorinated oil, and the perpendicular channel contains the dispersed phase, the saturated solution coming from the extraction zone (point **b** on figs.1 and S1). The syringe pumps control all the flows. Different fluorinated oils were tested (see Table S1), their viscosity varying by 2 orders of magnitude with roughly the same interfacial energy with water (almost the same chemical composition, except for FMS which contains fluoromethylsiloxan groups) and the same solvent compatibility. The most viscous, GPL106 Kryptox®, was selected because it permits good droplet stability (no coalescence – note that our set-up involves no surfactant) with temperature ranging from 5 to 80°C. To maintain the desired temperature in the generation module, all components were placed in an incubator allowing temperature control from room temperature to 80°C[19] (points **a** and **b** on figs.1 and S1).

*Droplet storage: generation of supersaturation, droplet Incubation and observation module*
Once saturated droplets are generated at a given temperature, tubes are sealed and stored in a water incubator (point **d** on figs.1 and S1) allowing temperature control from 5 to 80°C[20]. Solutions are generated at different concentrations in the plug factory of the set-up by switching from one outlet tube to another via a manual shut-off valve (point **c** on figs.1 and S1). They are stored in the thermostatted incubator 2 of the set-up at 80°C (at least always 10°C higher than extraction temperature) to prevent nucleation during storage and dissolve the few potential powder nanoparticles that went through the filter. This switch system is placed in an incubator at the same temperature as the plug factory to avoid unwanted crystallization during the passage between incubator 1 and 2 (fig.1). In all tubes, the concentration is the solubility at the temperature of generation (extraction). Then supersaturation is generated by applying a temperature variation profile to the droplet incubator. Time-sequence observation of droplets is carried out using a digital camera with various zooms (Opto GmbH) attached to a XYZ motorized arm[20].

*Bubble killer*
In our preliminary experiments, we noted the presence of bubbles at the filter outlet (point **a** on fig.1). These bubbles are due to solvent cavitation in the filter producing local depressions that trigger solvent evaporation. These solvent bubbles cause instabilities in the droplet generation (point **b** on fig.1). By moving in the P-T solvent phase diagram we compensate for these depressions by increasing pressure at the filter outlet so that solvent evaporation no longer occurs. Pressure is increased at constant flow rate by adding load losses downstream of the circuit. Here, adding 1m of 500µm ID tube at the output of the microfluidics set-up (point **e** on figs.1 and S1), makes it possible to multiply by approximately 3 the pressure at the outlet of the filter (according to the Poiseuille law) and kills bubbles.

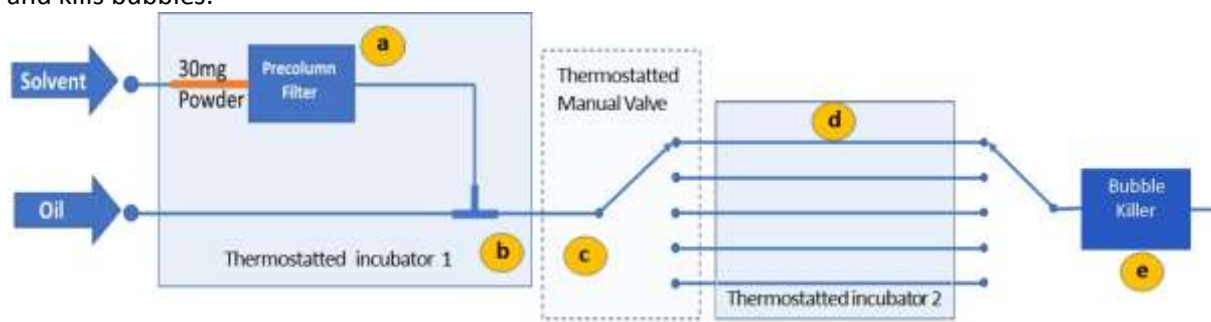

*Figure 1: Microfluidics set-up. (a) Microfluidics extraction, (b) Plug factory, (c) Thermostatted manual valve, (d) Droplet storage and (e) Bubble killer*

*Crystallization Procedure for solid form screening*
To test our microfluidics platform for solid phase screening using extremely small quantities, saturated droplets generated at 3 different temperatures (45, 55 and 70°C) are stored in 3 different sealed tubes at 80°C. The 3 solutions are at 3 different concentrations corresponding to solubilities at 45, 55 and 70°C (tables 1 and S2). Then the storage temperature is cooled to 10°C in steps of 5°C every 5 hours. Pictures of each droplet are taken automatically every 15 minutes to determine the nucleation temperature and observe possible transformations such as new nucleation events, solution-mediated-phase transition or crystal habit modification.

## Characterization of Solids

### In-situ Raman analysis

Sulfathiazole crystals are analyzed in situ at the end of the experiment using a Kaiser RXN1 Raman microscope system. Measurements are made at room temperature using a 785-nm laser at 10% power, with an exposure time of 5s and repeated 5 times. Tubes from the droplet storage zone are placed on the stage of the microscope and observed with a 10X objective lens. Droplets containing crystals are analyzed: the procedure is first, acquisition of a reference spectrum of a droplet containing solvent and Sulfathiazole in solution and second, acquisition of a droplet containing a crystal. This procedure allows us to identify the peaks that correspond to the Sulfathiazole crystal and compare them to reference spectra of sulfathiazole[21, 22] to identify polymorphs.

### X-Ray diffraction analysis

In addition to Raman analysis, X-Ray diffraction (XRD) analyses were performed using an XRD device mounted on a Cu rotating anode, Rigaku RU 200BH, with a 2D MAR 345 detector. Crystals were harvested from the PFA tubing using the procedure developed by Gerard et al.[20] and transferred in capillaries of 0.3 or 0.7 mm ID for XRD analysis. An exposure time of 360s was used.

## RESULTS AND DISCUSSION

## Solubility

Solubilities of Sulfathiazole (Form III) in water, isopropanol, and acetonitrile are shown in figure 2. To ensure that no phase transition occurs during the experiment, Raman spectra of the powder bed were taken before and after all experiments. Van't Hoff plot (fig.S2) and dissolution enthalpy values are given in Supplementary Information (table S3).

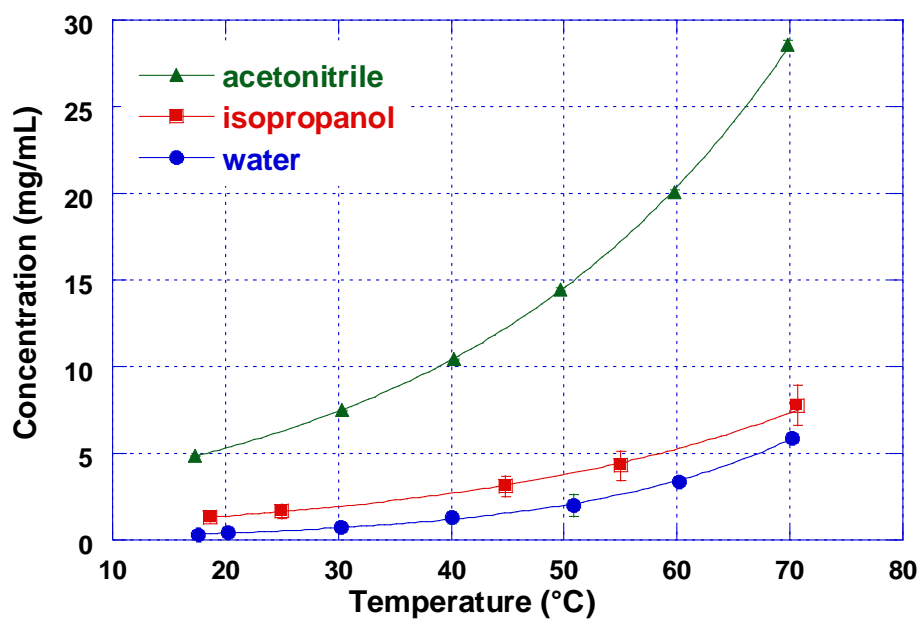

*Figure 2: Solubilities of Sulfathiazole form III in water, isopropanol, and acetonitrile. The curves are exponential fits to our measurements.*

## Sulfathiazole solid form screening

Because acetonitrile and isopropanol are better solvents for sulfathiazole than water, Sulfathiazole solubilities were greater in these solvents. However, nucleation was more difficult in these solvents (figs. S3 and S4). This result is in disagreement with the empirical rule that the higher the solubility, the easier nucleation[23]. In the literature, this phenomenon has already been observed, in particular by Anderson et al[24] who showed that nucleation kinetics in water are faster than in 1-propanol, ammonia and that there is no nucleation in nitromethane, although solubilities follow the reverse order. The difficulty of nucleation may therefore come from the solvent-solute association, as previously described by Gu et al[25] for the polymorphism of Sulfamerazine. Due to low nucleation of Sulfathiazole in isopropanol and acetonitrile, we discuss in the following only experiments in water.

### *Droplet observation and solid phase characterization by in-situ Raman spectroscopy*

We generated saturated droplets at 3 different temperatures (45, 55 and 70°C), corresponding to 3 initial concentrations $C_i$ (Table 1). These droplets are stored in 3 different sealed tubes at 80°C. After cooling from 80°C to 10°C, we have a set of about 5000 images at the end of the crystallization experiment. Figure 3 shows micrographs of representative droplets for each $C_i$. The different crystal habits obtained (fig.4), correspond to different polymorphs as confirmed by Raman and XRD characterization. From the time-sequence images produced during the experiment, we obtained nucleation statistics as plotted in figure 5 for each $C_i$.

| Temperature (°C) | $C_i$ (mg/mL) |
|---|---|
| 45 | 1.56 |
| 55 | 2.64 |
| 70 | 5.87 |

*Table 1: Concentration of Sulfathiazole in water at the different temperatures of preparation ($C_i$).*

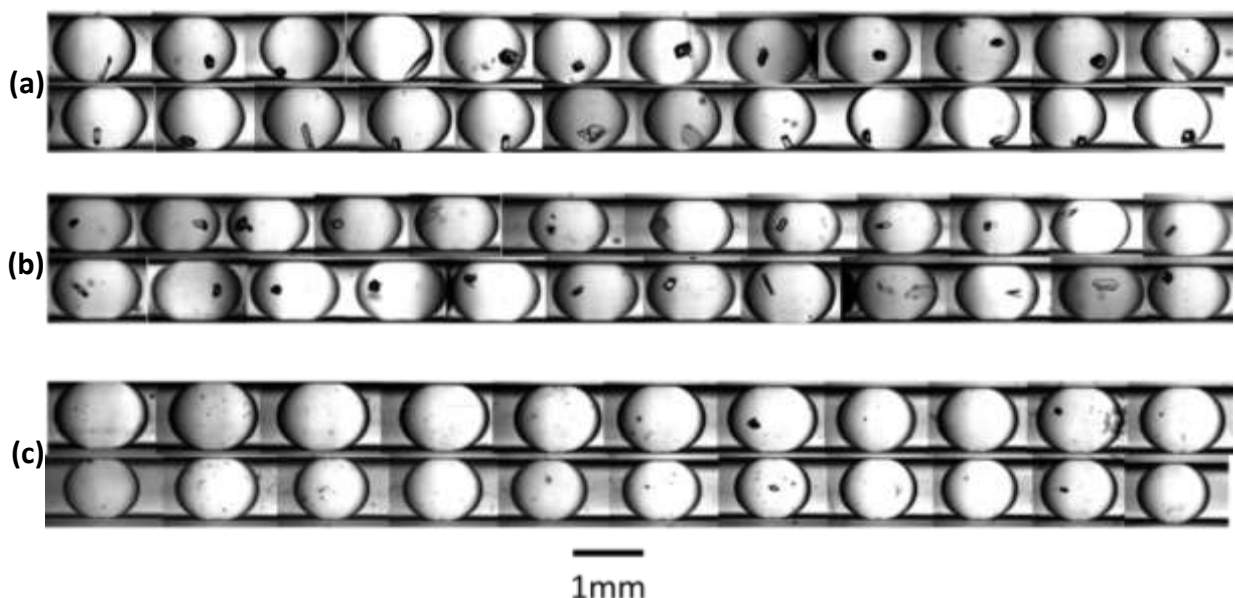

*Figure 3: Micrographs of droplets at the end of the experiment of crystallization of Sulfathiazole in water (a) droplets generated at 70°C ($C_i$=5.87mg/mL), (b) droplets generated at 55°C ($C_i$=2.64mg/mL), (c) droplets generated at 45°C ($C_i$=1.56mg/mL).*

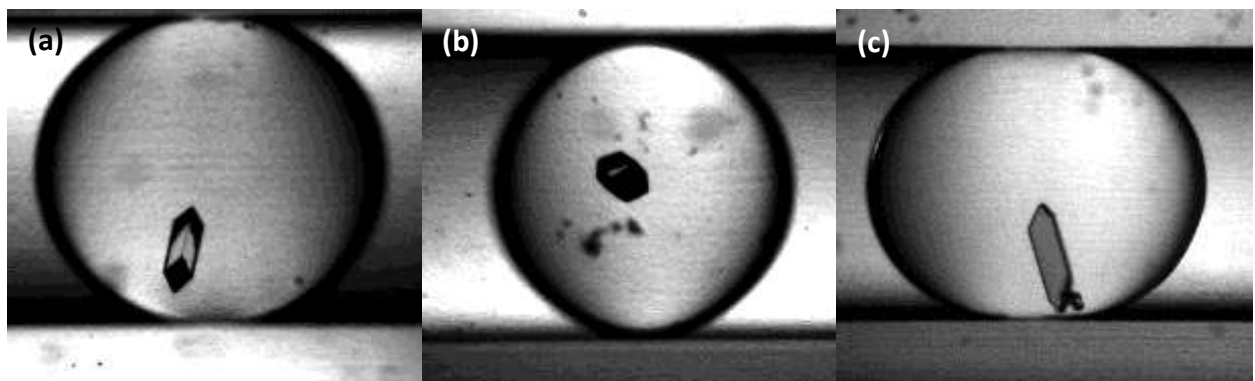

*Figure 4: Different crystal habits observed for crystallization of Sulfathiazole in water in 1mm droplets (a) form II, (b) form III and (c) form IV.*

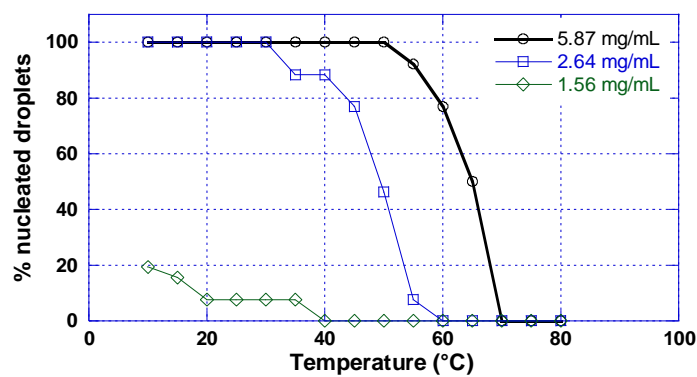

*Figure 5: Percentage of nucleated droplets of Sulfathiazole in water vs temperature during the cooling crystallization experiment, lines improve legibility.*

Crystals nucleated in about 60 droplets out of 90. They were all analyzed by Raman spectroscopy and typical spectra obtained are presented in figure 6. The sequential image acquisition allows us to confirm that there is not a second nucleation event nor a solution-mediated-phase transition in any droplet. This confirms that small volumes "freeze" (stabilize) the phase nucleated, as previously observed[26, 27]. Identification of the 3 polymorphs, II – III and IV, was confirmed by XRD on 3 samples, in agreement with the data from the Cambridge Structural Database files Suthaz, Suthaz02 and Suthaz04 respectively (https://www.ccdc.cam.ac.uk/).

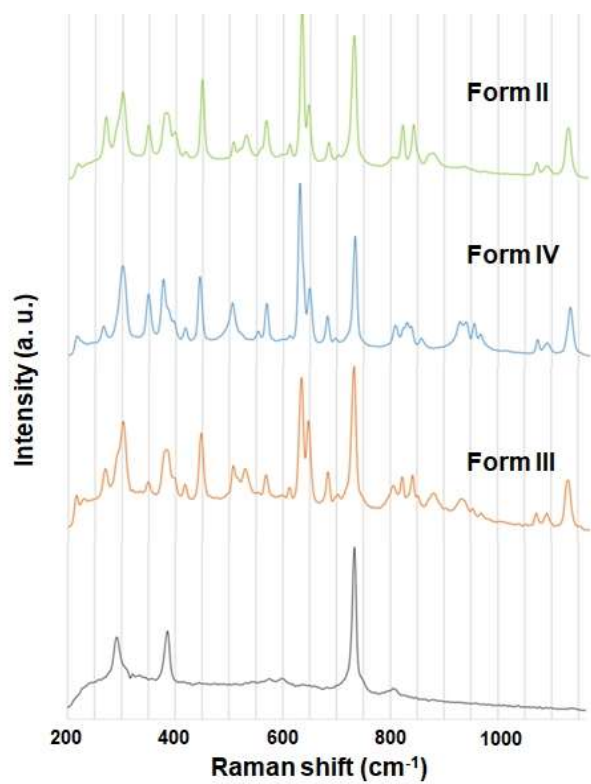

*Figure 6: Different Raman spectra observed for crystals of Sulfathiazole polymorphs in water in 1mm droplets. The bottom curve corresponds to a reference spectrum of a supersaturated droplet without crystals.*

The great advantage of this microfluidics approach lies in enabling numerous (here ~30) identical experiments to be performed, thereby ensuring consistent data on polymorph nucleation. The percentages of different polymorphs crystallized at the end of the experiments are summarized in table 2. Starting from a powder of form III, all the Ci yield polymorphs of forms II, III and IV. Ci of 2.64 and 5.87mg/mL lead mainly to form IV, i.e. in 69% of the droplets. A Ci of 1.56mg/mL leads only to 19.2% of nucleated droplets, a few droplets for each form. This highlight the advantage of generating a large number of identical crystallization experiments, notably due to stochasticity of nucleation.

|  | $C_i$=1.56 mg/mL | $C_i$=2.64 mg/mL | $C_i$=5.87 mg/mL |
|---|---|---|---|
| % Form II (# of crystallized droplets) | 40 (2) | 11.6 (3) | 19.2 (5) |
| % Form III (# of crystallized droplets) | 20 (1) | 19.2 (5) | 11.6 (3) |
| % Form IV (# of crystallized droplets) | 40 (2) | 69.2 (18) | 69.2 (18) |
| % of crystallized droplets | 19.2 | 100 | 100 |

*Table 2: percentages of different polymorphs crystallized at the end of the experiment*

Moreover, from figure 7, it is noteworthy that 2 crystals of form IV have nucleated at 55°C for a Ci of 2.64 mg/mL which is the solubility of form III. This can be due to the fact when temperature is cooled to 55°C in the thermostatted incubator 2, the temperature can pass by 1°C this temperature for a short time and then allows nucleation of a less stable phase. Or, the nucleation of the form IV at a concentration equal to the solubility of form III would confirm that form IV is the most stable phase as stated by Blagden et al.[18].

Figure 7 summarizes the data collected during the cooling crystallization experiment. In this screening, using 30mg of API, and only one cooling profile gives the 3 usual polymorphs of Sulfathiazole, forms II, III and IV. This information on the probability of nucleation of a given polymorph is useful in pharmaceutical development.

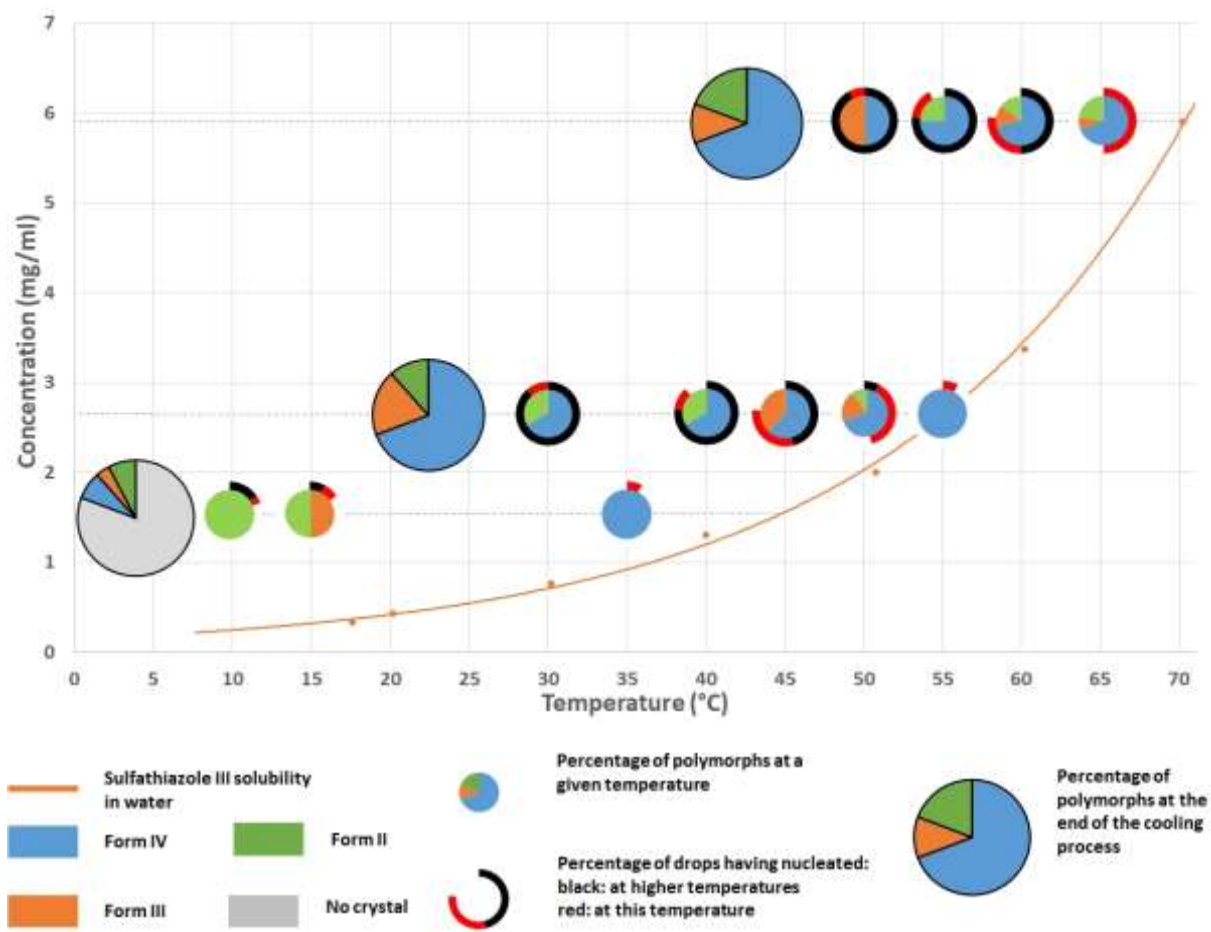

*Figure 7: Graphical summary of polymorphs obtained during cooling crystallization, to* 10°C by steps of 5°C every 5 hours, *of Sulfathiazole in water.*

## CONCLUSIONS

We present in this paper the development of a microfluidics platform for solid phase screening using extremely small quantities of raw materials. This microfluidics platform is based on our previous set-up for solubility measurement generating saturated solutions directly from powder. No solution in excess of that used for the droplet-crystallization experiment is required and the set-up is compatible with most solvents and molecules without using surfactant.

Using this microfluidics platform, we first measured the solubility of Sulfathiazole in water, isopropanol, and acetonitrile. Second, we performed a polymorph screening of Sulfathiazole using extremely small quantities of material, as little as 30 mg, for numerous identical cooling crystallization experiments from 80 to 10°C. In the experiments presented, we obtained the 3 usual polymorphs of Sulfathiazole. We show that this approach yields reliable information on the probability of nucleation of a given polymorph, useful in pharmaceutical development. Further studies could usefully explore different cooling policies that may influence the probability of nucleation of different polymorphs.

**ACKNOWLEDGMENTS:** *We thank Technologie Servier for financial support. We are grateful to T. Bactivelane (CINaM) and M. Audiffren (ANACRISMAT) for technical assistance. We thank Marjorie Sweetko for English revision.*

# Supplementary Information File

## Microfluidics platform for polymorph screening directly from powder

Guillem Peybernès[1,2], Romain Grossier[1], Frédéric Villard[2], Philippe Letellier[2], Nadine Candoni[1], Stéphane Veesler[1*],

[1]CNRS, Aix-Marseille University, CINaM (Centre Interdisciplinaire de Nanosciences de Marseille), Campus de Luminy, Case 913, F-13288 Marseille Cedex 09, France,
[2]Technologie Servier, 27 Rue EugèneVignat, 45000 Orléans, France
*veesler@cinam.univ-mrs.fr

Interfacial tensions between oil and water were measured by the pendant drop method using *Dataphysics OCA 20* (Optical Contact Angle) setup.

| Name | Density | Viscosity at 20°C (cSt) | Oil/water interfacial tension at 20°C (mN/m) |
|---|---|---|---|
| **FMS**[a] | 1,25 | 300-350 | 41 (±0.5) |
| **FC70**[a] | 1,94 | 12 (at 25°C) | 53.7 (±0.5) |
| **GPL103**[b] | 1.92 | 82 | 56.7 (±0.2) |
| **GPL105**[b] | 1.93 | 522 | 58.8 (±0.1) |
| **GPL106**[b] | 1.94 | 822 | 58.3 (±0.2) |

[a]: from Hampton Research and [b]: from DuPont

*Table S1 : Typical properties of FMS, FC70 Fluorinert® and Krytox® fluorinated oils*

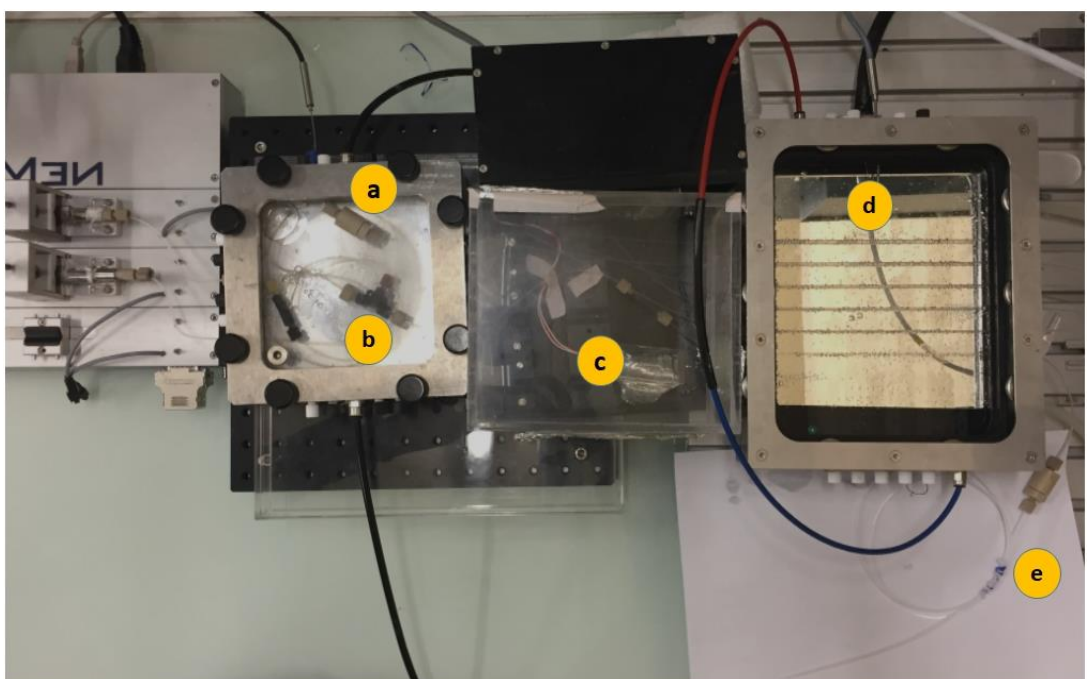

*Figure S1: Microfluidics set-up. (a) Microfluidics extraction, (b) Plug factory, (c) Thermostatted manual valve, (d) Droplet storage and (e) Bubble killer*

| Temperature (°C) | Concentration (mg/mL of water) | Concentration (mg/mL of isopropanol) | Concentration (mg/mL of acetonitrile) |
|---|---|---|---|
| 45 | 1.56 | 3.20 | 12.30 |
| 55 | 2.64 | 4.46 | 17.22 |
| 70 | 5.87 | 7.35 | 28.52 |

*Table S2: Concentration of sulfathiazole in water, isopropanol and acetonitrile at the different preparation temperatures*

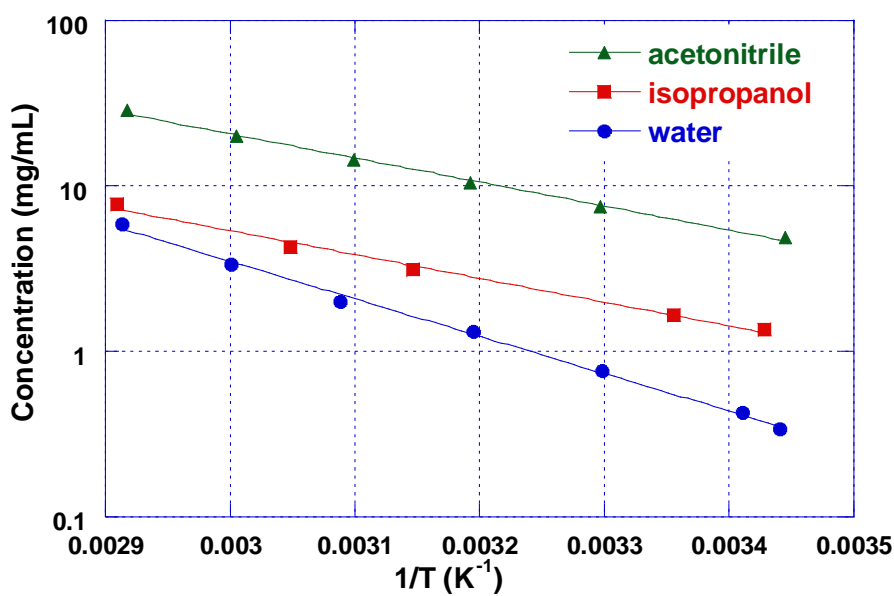

*Figure S2 : van't Hoff plot of sulfathiazole form III solubilities in different solvents*

| Solvent | $\Delta H_d$ (kJ/mol) |
|---|---|
| acetonitrile | 27.8 |
| isopropanol | 27.5 |
| water | 43.3 |

*Table S3: Dissolution Enthalpies in different solvents from van't Hoff plot of sulfathiazole form III solubilities (fig.S1)*

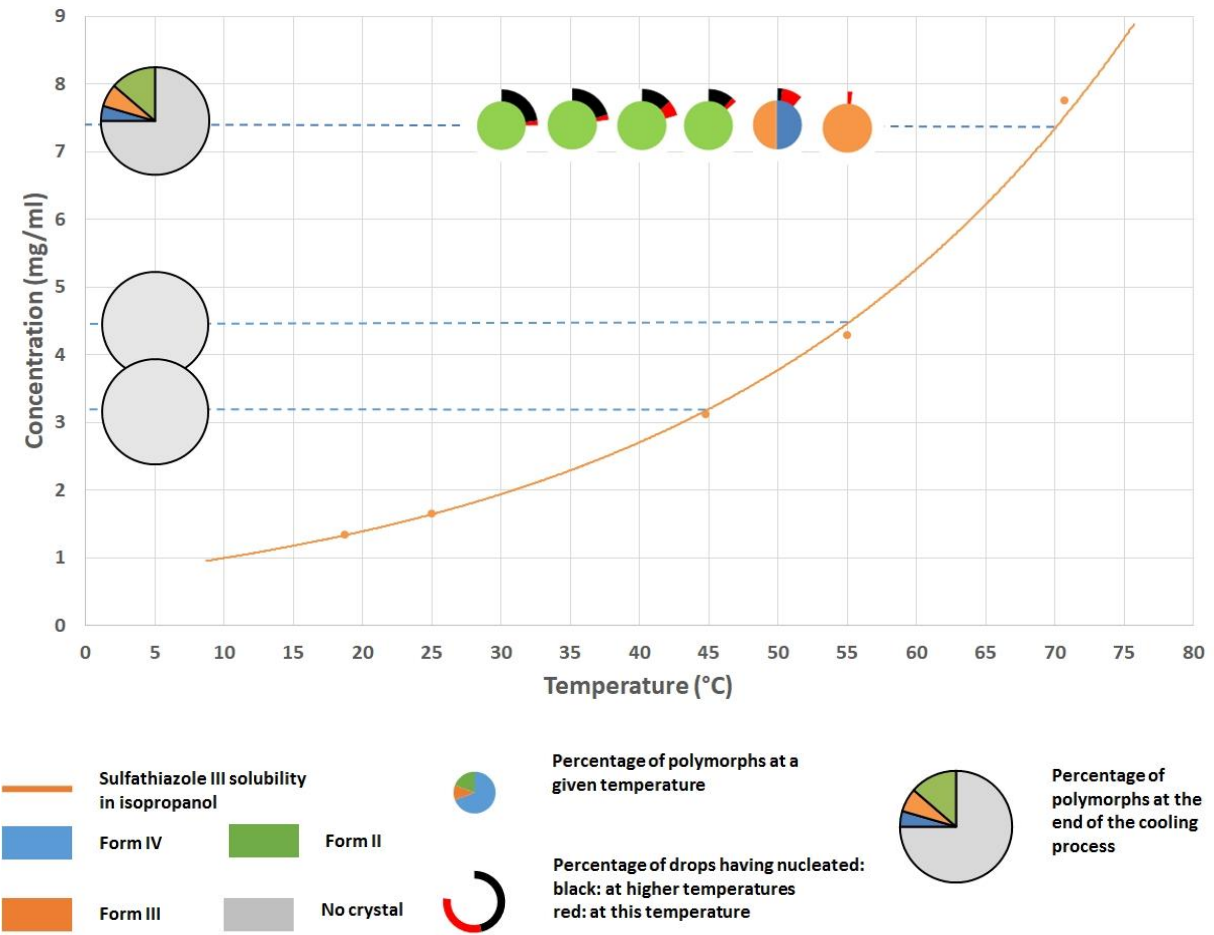

*Figure S3: Graphical summary of polymorphs obtained during cooling crystallization, to* 10°C by steps of 5°C every 5 hours, *of Sulfathiazole in isopropanol.*

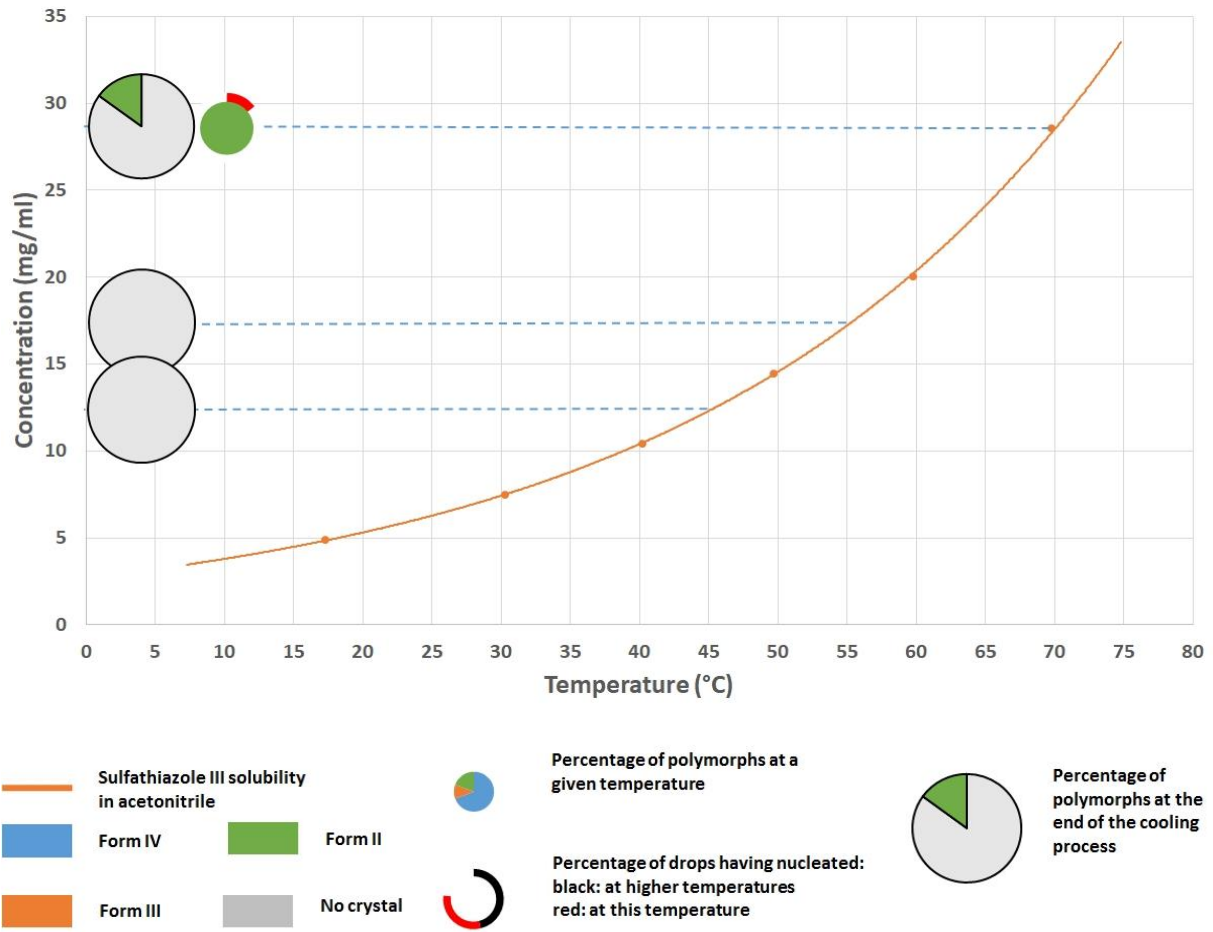

*Figure S4: Graphical summary of polymorphs obtained during cooling crystallization, to* 10°C by steps of 5°C every 5 hours, *of Sulfathiazole in acetonitrile.*